\documentstyle{article}
\input epsf
\begin{document}
\title{Approximations of the self-similar solution for blastwave \\
	in a medium with power-law density variation}
\author{O.~Petruk \\
	Institute for Applied Problems in Mechanics and Mathematics \\
       	NAS of Ukraine, 3-b Naukova St., Lviv 79000, Ukraine \\
        petruk@astro.franko.lviv.ua}
\maketitle 


\begin{abstract}
Approximations of the Sedov self-similar solution for a strong point 
explosion in a medium with the power-law density distribution 
$\rho^o\propto r^{-m}$ are reviewed and their accuracy are analyzed. 
Taylor approximation is extended to cases $m\neq 0$. 
Two approximations of the solution 
are presented in the Lagrangian coordinates for spherical, cylindrical and 
plane geometry. These approximations may be used for the investigation of 
the ionization structure of the adiabatic flow, i.e., inside 
adiabatic supernova remnants. 
\end{abstract}

\section {Introduction}

Self-similar (Sedov \cite{Sedov-1946b,Sedov}) solutions for the strong 
point explosion 
in the uniform medium $\tilde{\rho}^{o}={\rm const}$ or in a medium with 
power-law density distribution 
\begin{equation}
\label{rho-power}
\tilde{\rho}^{o}(\tilde{r})=\tilde{\rho}^{o}(0){r}^{-m},
\end{equation}
where $\tilde{r}$ is the distance from the center of explosion, are 
widely used for modelling the adiabatic supernova remnants, solar flares 
and processes in active galactic nuclei. \par

Sedov (\cite{Sedov-1946b,Sedov}) has obtained the exact solution 
solving the system of hydrodynamic differential equation on the base 
of dimensional methods. 
Independently, Taylor (\cite{Taylor-50}) has solved the 
same task in the case of the uniform medium numerically and 
in analitical form approximately.  The main Taylor's idea was 
to approximate the fluid velocity variation behind the shock front. \par

Kahn (\cite{Kahn}) have proposed the approximation of the Sedov 
solution in the uniform medium. 
His technic consists in approximation of mass distribution inside the 
shock\-ed region. 
Using Kahn methodology, Cox \& Franco (\cite{Cox_Fanko-81}) have built the 
approximation of the exact solution in the power-law medium 
(\ref{rho-power}) for $m<2$. 
With the same technic, Cox \& Anderson (\cite{Cox-And}) have presented the 
approximation for description of the shocked region and blastwave motion 
in uniform medium of finite pressure. \par

Ostriker \& McKee (\cite{Ostriker-McKee-88}) basing on the virial 
theorem have given a number of approximations for 
the fluid characteristic variation as one- or two-power polinoms. \par

Hnatyk (\cite{Hn87}) have proposed to approximate firstly the relation between 
the Eulerian and Lagrangian coordinates of the flow elements. \par

In present work, Taylor approximation is written for the medium with 
power-law density variation (\ref{rho-power}). 
Using Hnatyk's approach, we develope also two 
approximations of the Sedov solution for power-law medium with $m\le 2$ in 
Lagrangian coordinates 
that is useful for investigations of the nonequilibrium ionization 
processes in a shocked plasma, e.g., inside the adiabatic supernova 
remnants. One of the approximations presented here bases on 
the approximate hydrodynamic method for description of the 
nonspherical strong point explosion in the medium with arbitrary 
large-scale nonuniformity developed by Hnatyk \& Petruk (\cite{Hn-Pet-99}). 
Therefore, it may also be considered as additional test on this method. \par 

\section{Sedov solution and its approximations}

\subsection{Sedov solution}

If strong ($P_{\rm s}/P^{o}_{\rm s}\rightarrow\infty$) point 
($R_{o}/R\rightarrow0$) explosion with finite 
energy $E_{o}$ becomes in the point with coordinate $\tilde{r}=0$ 
in time $t=0$, the blastwave creates and propagates with velocity $D$ 
in the ambient medium with density $\tilde{\rho}^{o}(\tilde{r})$ 
($P_{\rm s}$ and $P^{o}_{\rm s}$ are pressure of the shocked gas 
and the gas of the ambient medium at the shock 
front position, $R_{o}$ is the size of the body exploded, $R$ 
is the radius of the blastwave). It is also assumed that 
injected mass is small and  
no energy lost from the shocked region during the motion. \par

Such a task is described by a system of hydrodynamic differential 
equations. 
Sedov (\cite{Sedov}) gives the analytical self-similar 
solution for description of the motion of shock front and the 
distribution of fluid parameters inside the shocked region for a 
strong point explosion in the uniform ambient medium and in the center 
of symmetry of a radially stratified medium (\ref{rho-power}). \par

This solution shows that the strong blastwave in a medium with the 
power-law density distribution (\ref{rho-power}) moves with 
deceleration then $m<N+1$ and accelerates then $m>N+1$. 
If $m\geq N+1$, both the mass inside any sphere, 
which containes the center of the symmetry, and kinetic energy 
equal infinity. We will consider $m<N+1$ cases only. \par

Radius $R$ and velocity $D$ of the strong blastwave in the medium 
(\ref{rho-power}) with $m<N+1$ are (Sedov \cite{Sedov}): 
\begin{equation} 
\label{R_s_Sedov-ro^w} 
R=\left({E_o\over\alpha_A\ \tilde{\rho}^o(0)}\right)^{1/(N+3-m)}
t^{\ 2/(N+3-m)}\ , 
\end{equation}
\begin{equation} 
\label{D_Sedov-ro^w} 
D(R)={2\over N+3-m}\left({E_o\over\alpha_A\ \tilde{\rho}^o(0)} 
\right)^{1/2} R^{-(N+1-m)/2}\ ,
\end{equation} 
\noindent
where $N=0,1,2$ for plane, cylindrical and spherical wave, respectively, 
$\alpha_A$ is a self-similar constant. \par

Distributions of the fluid characteristics behind the shock front 
are self-similar, i.e., for any time $t$ the density $\tilde{\rho}$, 
pressure $\tilde{P}$, 
fluid velocity $\tilde{u}$ variations and coordinate $\tilde{a}$ are 
\begin{equation}
\label{uni-prof-1}
\tilde{\rho}(\tilde{r},t)=\tilde{\rho}_{\rm s}(t)\cdot\rho(r),
\end{equation}
\begin{equation}
\label{uni-prof-2}
\tilde{P}(\tilde{r},t)=\tilde{P}_{\rm s}(t)\cdot P(r),
\end{equation}
\begin{equation}
\label{uni-prof-3}
\tilde{u}(\tilde{r},t)=\tilde{u}_{\rm s}(t)\cdot u(r),
\end{equation}
\begin{equation}
\label{uni-prof-4}
\tilde{a}(\tilde{r},t)=R(t)\cdot a(r),
\end{equation}
where $r=\tilde{r}/R(t)$, $\tilde{a}$ is the original position of the 
fluid mass element and superscript "s" corresponds to values of the 
parameters at the shock front (Fig.~\ref{accuracy_Taylor_Kahn}). \par

Gas occupies whole shocked region ($0\le \tilde{r}\le R$) when $m\le m_1$, 
\begin{equation} 
m_1={1+3N+(1-N)\gamma\over\gamma+1}\ . 
\end{equation}
When $m\rightarrow m_1$ central pressure $P(0)\rightarrow 0$. 
Shock waves in media with steep density gradients ($m>m_1$) develop 
a cavity around the center of explosion. 
Such a cavity creates in the uniform medium ($m=0$) when 
$\gamma>\gamma_1=(1+3N)/(N-1)$. 
Sedov has also presented a solution for hollow blastwaves. 
Review of approximations for these cases is given by Ostriker \& McKee 
(\cite{Ostriker-McKee-88}). We do not consider $m>m_1$ in this paper. \par 

For $m=m_1$ (or $\gamma=\gamma^*$ in the uniform medium) solution has very 
simple form: 
\begin{equation} 
\begin{array}{l}
\rho(r)=r^{N-1}\ ,\qquad P(r)=r^{N+1}\ , \\ \\
u(r)=r\ ,\hspace{1.4cm} a(r)=r^{(\gamma+1)/(\gamma-1)}\ . 
\end{array}
\label{solut-m_1}
\end{equation}

Singularities in the solution also appear with 
$m_2=(N+1)(2-\gamma)$ and 
$m_3=(2\gamma+N-1)/\gamma$ 
then some exponents in the solution equal infinity. 
Similarity solutions for these cases are deduced by 
Korobejnikov \& Rjazanov (\cite{Korobejnikov-Rjazanov-59}). 
For $N=2$ and $\gamma=5/3$ $m_1=2$, $m_2=1$, $m_3=13/5$. \par

Self-similar constant $\alpha_A=\alpha_A(N,\gamma,m)$ in equations 
(\ref{R_s_Sedov-ro^w})-(\ref{D_Sedov-ro^w}) 
for $R$ and $D$ may be found 
from the energy balance equation with variations of density $\tilde{\rho}$, 
pressure $\tilde{P}$ and mass velocity $\tilde{u}$ inside the shocked region 
\begin{equation} 
\label{enery-bal} 
{E_o\over \sigma}=\int\limits_0^R {\tilde{\rho}(r,t)
\tilde{u}(r,t)^2\over 2} r^Ndr+
\int\limits_0^R {\tilde{P}(r,t)\over \gamma-1} r^Ndr\ ,
\end{equation}
where 
$\sigma=4\pi$ for $N=2$, $\sigma=2\pi$ for $N=1$ and $\sigma=2$ for $N=0$ 
or, generally, 
$\sigma=2\pi N+(N-1)(N-2)$. 
If we proceed to normalized parameters using 
(\ref{uni-prof-1})-(\ref{uni-prof-3}) and 
general shock front conditions 
\begin{equation}
\tilde{\rho}_{\rm s}={\gamma+1\over\gamma-1}\tilde{\rho}^o_{\rm s},\ 
\tilde{P}_{\rm s}={2\over\gamma+1}\tilde{\rho}^o_{\rm s}D^2,\ 
\tilde{u}_{\rm s}={2\over\gamma+1}D
\end{equation}
we will obtain that $E_o=\beta_A\cdot MD^2/2$ with 
\begin{equation}
M=\sigma\tilde{\rho}^o(0)R^{N+1-m}/(N+1-m), 
\end{equation}
constant shape-factor 
\begin{equation}
\beta_A={4(N+1-m)\over\gamma^2-1}\cdot\left(I_{\rm K}+I_{\rm T}\right)
\label{beta_A} 
\end{equation}
and constant integrals 
\begin{equation}
I_{\rm K}=\int\limits_0^1 \rho(r)u(r)^2 r^Ndr\ , \qquad 
I_{\rm T}=\int\limits_0^1 P(r) r^Ndr\ . 
\end{equation}
Also we will have a self-similar constant 
\begin{equation} 
\label{alpha_A-vs-beta_A} 
\alpha_A={2\sigma\over(N+1-m)(N+3-m)^2}\cdot\beta_A\ .
\end{equation}
Simple formula gives $\alpha_A(N,\gamma,m_1)$: 
\begin{equation}
\alpha_A={2\sigma(\gamma+1)\over(N+1)(\gamma-1)\big((N+1)\gamma-N+1\big)^2}\ .
\end{equation}

The distributions (\ref{uni-prof-1})-(\ref{uni-prof-4}) 
in the exact solution are parametric 
functions of an internal parameter. The expressions for the functions are 
complicated. These factors stimulate developing the approximations of the 
self-similar solution. \par

\subsection{Taylor approximation}

Basing on own numerical results, Taylor (\cite{Taylor-50}) propose to 
approximate the velocity variation $u(r)$ behind spherical ($N=2$) shock 
front moving into the uniform medium ($m=0$) as 
\begin{equation} 
{\tilde{u}(r,t)\over D}={r\over\gamma}+\alpha r^n,
\label{app-Taylor-base}
\end{equation} 
where $\alpha$ and $n$ are found to give exact values of $\tilde{u}_{\rm s}$, 
$\tilde{P}_{\rm s}$, $\tilde{\rho}_{\rm s}$ and their first derivatives in 
respect to $r$. 
Substituting this approximation into the continuity equation and into the 
equation of state for perfect gas, the approximated distributions of the 
density and pressure obtain. 
Taylor do not give the dependence $a(r)$, but it may be taken from the 
adiabaticity condition $P(a)\rho(a)^{-\gamma}=P(r)\rho(r)^{-\gamma}$ 
and (\ref{appr_3a})-(\ref{appr_3b}): 
\begin{equation} 
a^{\gamma m-(N+1)}=P(r)\rho(r)^{-\gamma},
\label{a-vs-Prho}
\end{equation} 
with approximations for $P(r)$ and $\rho(r)$. \par

\begin{figure}
\epsfxsize=8.8truecm
\centerline{\epsfbox{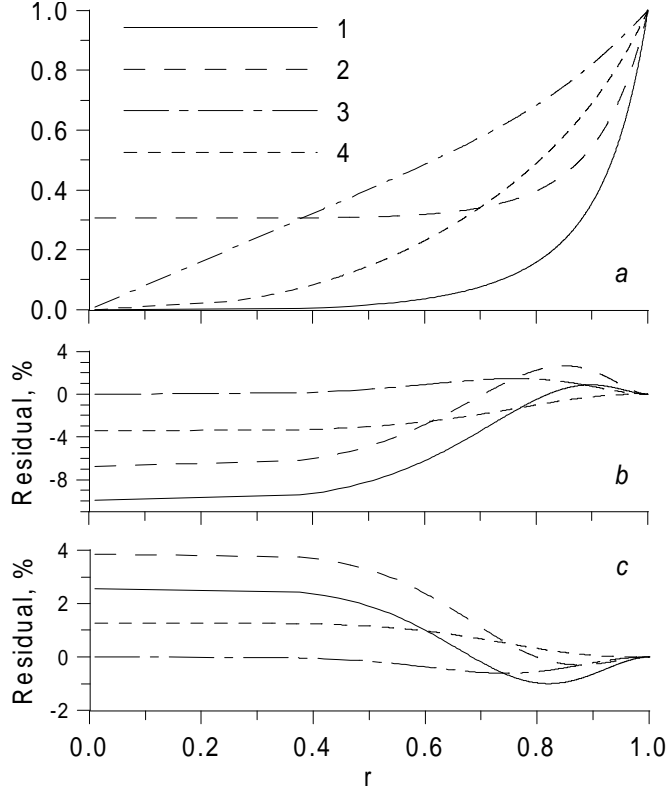}}
\caption[]{{\bf a-c.} Sedov solution and the accuracy of Taylor and Kahn 
approximations of the solution in the uniform medium: 
{\bf a}~exact Sedov solution, 
{\bf b}~relative differences of Taylor approximation, 
{\bf c}~relative differences of Kahn approximation. 
Lines: 1 -- $\rho(r)$, 2 -- $P(r)$, 3 -- $u(r)$, 4 -- $a(r)$. $\gamma=5/3$.
           }
\label{accuracy_Taylor_Kahn}
\end{figure}

So, Taylor approximation for the variations of density $\rho$, pressure $P$, 
fluid velocity $u$ and coordinate $a$ are: 
\begin{equation} 
\label{Taylor-n} 
\rho(r)={\rho(r,t)\over \rho_{\rm s}(t)}=
r^{\ 3/(\gamma-1)}\ \left({\gamma+1\over\gamma}-{r^{n-1}\over\gamma}\right)^{-p},
\end{equation}
\begin{equation} 
\label{Taylor-T} 
P(r)={P(r,t)\over P_{\rm s}(t)}=
\left({\gamma+1\over\gamma}-{r^{n-1}\over\gamma}\right)^{-q},
\end{equation}
\begin{equation} 
\label{Taylor-u} 
u(r)={u(r,t)\over u_{\rm s}(t)}=
{\gamma+1\over 2}\left({r\over\gamma}+{\gamma-1\over\gamma+1}{r^{n}\over\gamma}\right),
\end{equation}
\begin{equation} 
\label{Taylor-a} 
a(r)={a_o(r)\over R(t)}=
r^{\ \gamma/(\gamma-1)}\ \left({\gamma+1\over\gamma}-{r^{n-1}\over\gamma}\right)^{-s},
\end{equation}
where $n=(7\gamma-1)/(\gamma^2-1)$, $p=2(\gamma+5)/(7-\gamma)$, 
$q=(2\gamma^2+7\gamma-3)/(7-\gamma)$, $s=(\gamma+1)/(7-\gamma)$. 
Self-similar constant $\alpha_A=\alpha_A(2,\gamma,0)$ goes with 
(\ref{alpha_A-vs-beta_A}) and approximated profiles of $\rho$, $P$ 
and $u$. 
Fig.~\ref{accuracy_Taylor_Kahn} and table \ref{alpha_comp} demonstrate 
accuracy of Taylor approximation in comparison with the exact solution. \par

This approximation is extended to cases $m\neq 0$ in section 
\ref{Taylor-ext}. \par

\subsection{Kahn approximation}

Kahn (\cite{Kahn}) apply his methodology to the strong spherical blastwave 
($N=2$) in uniform medium ($m=0$) with $\gamma=5/3$. 
It is proposed to approximate first the mass distribution 
\begin{equation} 
\label{mu-def} 
\mu(r)={M(r,t)\over M_{\rm s}(t)}=
3\int\limits^r_0 \rho(r)r^2dr.
\end{equation}

Sedov solution shows that $P_r(r)=0$ near the centre 
(subscript "$r$" denotes a partial derivative in respect to $r$). 
This fact allows to find that $\mu_r/\mu=15/(2r)$ at $r=0$.  
On the base of the equation of motion,  
$\mu_r/\mu=12$, $\mu_{rr}=168$ and $(\mu_r/\mu)_r=24$ at $r=1$. 
Therefore ratio $\mu_r/\mu$ is proposed to be approximated as 
\begin{equation} 
\label{Kahn-dmu-appr} 
{\mu_r\over\mu}=
{15\over 2r}+{9\over 2}r^7.
\end{equation}
This formula satisfies all written boundary conditions at both ends. \par

The mass distribution finds as integral from (\ref{Kahn-dmu-appr}): 
\begin{equation} 
\label{Kahn-mu-appr} 
\mu(r)=r^{15/2}\ \exp\left({9\over16}\left(r^8-1\right)\right).
\end{equation}
Density distribution follows from (\ref{mu-def}) and (\ref{Kahn-mu-appr}): 
\begin{equation} 
\rho(r)=\mu_r/3r^2.
\end{equation}
Adiabaticity condition gives pressure variation 
\begin{equation} 
P(r)=\left({2\over3}\right)^{5/3}{1\over 32}\ {\mu_r^{5/3}\over \mu\ r^{10/3}}\ . 
\end{equation}
Velocity deduses from the mass conservation equation 
\begin{equation} 
u(r)={4\over 3}r-{4\mu\over\mu_r}\ .
\end{equation}
If present location of mass element $a$ is $r$, then a(r) may be 
found from the condition of mass conservation 
$\mu(a)=\mu(r)$ and relation (\ref{mu(a)}) $\mu(a)=a^{3}$: 
\begin{equation} 
\label{Kahn-a-appr} 
a(r)=\mu(r)^{1/3}\ .
\end{equation}

The expressions for Kahn approximation are the same as 
(\ref{Kahn-n})-(\ref{Kahn-mu}) with $m=0$. 
The accuracy of this approximation are shown on the 
Fig.~\ref{accuracy_Taylor_Kahn}. 

\subsection{Approximation of Cox \& Franco}

Appling Kahn's approximation technique, Cox \& Franco 
(\cite{Cox_Fanko-81}) obtain the approximation of the 
self-similar solution for an ambient medium with the power-law density 
distribution (\ref{rho-power}) with $m<2$ for $\gamma=5/3$ and $N=2$. 
Approximation of Cox \& Franco are: 
\begin{equation} 
\begin{array}{l} 
{\displaystyle 
\rho(r)=
\left({5\over8}+{3\over8}r^{\ 8-4m}\right)\cdot r^{\ (9-5m)/2} 
}\\ \\ 
{\displaystyle \qquad\qquad\qquad\!\!
\times\exp\left({3\over8}{(3-m)\over(2-m)}\left(r^{\ 8-4m}-1\right)\right)\ , 
}
\end{array} 
\label{Kahn-n} 
\end{equation}
\begin{equation} 
\begin{array}{l} 
{\displaystyle 
P(r)=
\left({5\over8}+{3\over8}r^{\ 8-4m}\right)^{5/3}
}\\ \\ 
{\displaystyle \qquad\qquad\qquad\!\!
\times\exp\left({3\over4(2-m)}\left(r^{\ 8-4m}-1\right)\right)\ , 
}
\end{array} 
\label{Kahn-T} 
\end{equation}
\begin{equation} 
\label{Kahn-u} 
u(r)=
4\ r\ {1+r^{\ 8-4m}\over 5+3\ r^{\ 8-4m}}\ ,
\end{equation}
\begin{equation} 
{\displaystyle 
a(r)=
r^{5/2}\ \exp\left({3\over8(2-m)}\left(r^{\ 8-4m}-1\right)\right)\ ,}
\label{Kahn-a} 
\end{equation}
\begin{equation} 
{\displaystyle 
\mu(r)=r^{\ 5(3-m)/2} 
\ \exp\left({3\over8}{(3-m)\over(2-m)}\left(r^{\ 8-4m}-1\right)\right)\ . }
\label{Kahn-mu} 
\end{equation}
Author's approximation for $\beta_A$ is 
\begin{equation} 
\label{beta-Cox-Fanko} 
\beta_A=1.125\cdot(0.22+0.52\cdot(3-m)/3).
\end{equation}

\begin{figure}
\epsfxsize=8.8truecm
\centerline{\epsfbox{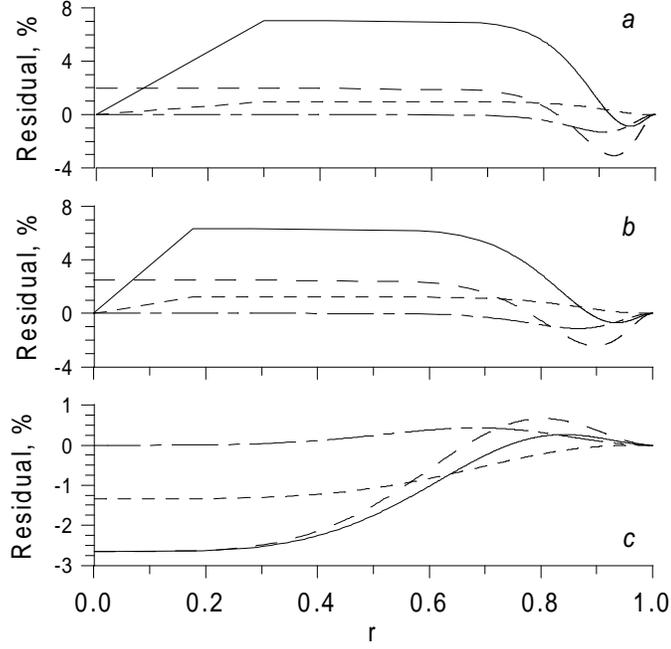}}
\caption[]{{\bf a-c.} Accuracy of Cox \& Franco approximation of the 
self-similar solution in the power-law medium (\ref{rho-power}):  
{\bf a}~relative differences of the approximation for $m=-4$, 
{\bf b}~relative differences for $m=-2$, 
{\bf c}~relative differences for $m=1$. 
Lines are the same as on Fig.~\ref{accuracy_Taylor_Kahn}. 
Fracture in the curves for $\rho(r)$ and $a(r)$ 
is due to very strong dependence of the relevant Sedov distributions 
on the internal parameter, which changes in these wide intervals 
of $r$ on $10^{-10}$ only. 
           }
\label{accuracy_Kahn_N-2}
\end{figure}

The accuracy of Cox \& Franco approximation is shown on 
Fig.~\ref{accuracy_Kahn_N-2} and table \ref{alpha_comp}. \par

\subsection{Approximations of Ostriker \& McKee}
\label{Ostr-McKee}

Ostriker \& McKee (\cite{Ostriker-McKee-88}) in the frame of the virial 
theorem approach applied to spherical blastwave (N=2) in 
the power-law ambient medium (\ref{rho-power}) 
and time-dependent energy injection $E_o(t)\propto t^{s}$, present a 
number of approximations for the self-similar solution. We consider further 
$s=0$. \par 

Authors introduce the dimensionless moments of coordinate $r$ and velocity 
$u$: 
\begin{equation} 
\label{moment-Ostr-McKee} 
K_{ij}=l_\mu\int\limits_0^1 r^iu(r)^j\rho(r)r^2dr\ ,
\end{equation} 
where $l_\mu=(\gamma+1)(3-m)/(\gamma-1)$, 
and consider three types of approximations for $u(r)$ and $\rho(r)$: 
linear velocity approximation (LVA)
\begin{equation} 
\label{LVA-Ostr-McKee} 
u(r)=r,\qquad \rho(r)=r^{(6-(\gamma+1)m)/(\gamma-1)},
\end{equation} 
one-power aproximation (OPA) 
\begin{equation} 
\label{OPA-Ostr-McKee} 
u(r)=r^{l_u},\qquad \rho(r)=r^{l_\rho},
\end{equation} 
and two-power aproximation (TPA) 
\begin{equation} 
\label{TPA-Ostr-McKee-u} 
u(r)=a_ur^{l_{u,1}}+(1-a_u)r^{l_{u,2}},\\ \\
\end{equation} 
\begin{equation} 
\label{TPA-Ostr-McKee-rho} 
\rho(r)=a_\rho r^{l_{\rho,1}}+(1-a_\rho)r^{l_{\rho,2}}.
\end{equation} 

In such an approach the self-similar constant $\alpha_A$ as well as 
exponents $l_u$ and $l_\rho$ may be expressed in terms of moments $K_{02}$ 
and $K_{11}$. 
Namely, under self-similarity $\alpha_A=2\pi\eta^2\beta_A/(3-m)$, where 
$\eta=2/(5-m)$ and factor $\beta_A$ equals 
\begin{equation} 
\label{beta_A-Ostr-McKee} 
\beta_A={2\over3}\cdot{2K_{02}(3\gamma-5)+(5-m)(\gamma+1)K_{11}
\over (\gamma^2-1)(\gamma+1)}\ .
\end{equation} 
Exponents in OPA are 
\begin{equation} 
l_u={2K_{20}-K_{11}(1+K_{20})\over(1-K_{20})K_{11}}\ , \quad 
l_{\rho}={5K_{20}-3\over1-K_{20}}\ .
\end{equation} 

Derivatives at shock front are used to obtain the moments. So, 
\begin{equation} 
K_{ij}={1\over 1+s_{ij}/l_\mu}\ ,
\label{K_Ostr-McKee}
\end{equation} 
where $s_{ij}=i+j$ in LVA and $s_{ij}=i+j+j(m_1-m)/2$ in OPA. 
Using (\ref{K_Ostr-McKee}) $\alpha_A$ may be written in a simple 
form in LVA: 
\begin{equation} 
\label{alpha_A-Ostr-McKee} 
\alpha_A={16\pi\over 3(5-m)^2}\cdot
{11\gamma-5-m(\gamma+1)\over 
(\gamma^2-1)\big(5\gamma+1-m(\gamma+1)\big)}\ .
\end{equation}

Moments have more complicated form in TPA. 
In this approach the expression for $u(r)$ coinsides with the approximation 
(\ref{Taylor-u}) of Taylor with $n=1+\gamma(m_1-m)/(\gamma-1)$ that 
equals to Taylor's $n$ at $m=0$. So, TPA is extension of Taylor 
approximation of $u(r)$ to $m\neq 0$. Contrary to Taylor's approach to find 
$\rho(r)$ and $P(r)$ from the hydrodynamic equations, Ostriker \& McKee find 
the density variation 
independently as TPA (\ref{TPA-Ostr-McKee-rho}) with 
\begin{equation} 
\begin{array}{l}
{\displaystyle a_{\rho}={\gamma(m_1-m)\over 10-\gamma-(\gamma+2)m} \ ,}\\ \\
{\displaystyle l_{\rho,1}={3-\gamma m\over\gamma-1} \ ,
\quad l_{\rho,2}={6+(\gamma+1)(m_1-2m)\over \gamma-1} \ ,}
\end{array}
\end{equation} 
where $\gamma>1$. 
For $m=0$ variation $\rho(r)$ in TPA coinsides with the result of Gaffet 
(\cite{Gaffet}) for case of uniform medium (Ostriker \& McKee 
\cite{Ostriker-McKee-88}). Two-power velocity approximation is used to 
extend Taylor approximation to cases $m\neq 0$ in section \ref{Taylor-ext}. \par

Pressure distribution are also restored independently. 
It may be found in OPA as a linear pressure approximation 
(LPA) and for TPA in the frame of pressure-gradient approximation (PGA). \par

Most of mass is concentrated near the shock front and distribution 
$u(r)$ is close to a linear function of $r$. Therefore, as noted by 
Gaffet (\cite{Gaffet}), the right side of Euler equation 
\begin{equation} 
{\partial\tilde{P}(\tilde{r},t)\over \partial M(\tilde{r},t)}=
-{1\over4\pi}{1\over \tilde{r}^2}{d\tilde{u}(\tilde{r},t)\over dt}
\label{Eul-eq}
\end{equation} 
is nearly a constant. LPA (Gaffet \cite{Gaffet}, Ostriker \& McKee 
\cite{Ostriker-McKee-88}) use this feature assuming the 
pressure to be a linear function of the mass fraction $\mu(r)$
\begin{equation} 
P(r)=P(0)+(P^*_{\rm s}/l_{\mu})\ \mu(r).
\end{equation} 
Logariphmic derivative of pressure at the shock front is 
$P^*_{\rm s}=(d\ln P/d\ln r)_{\rm s}=
(2\gamma^2+7\gamma-3-\gamma m(\gamma+1))/(\gamma^2-1)$. 
Mass in OPA is $\mu(r)=3l_\mu^{-1}r^{l_\mu}$. 
P(0) in LPA is (Gaffet \cite{Gaffet}) 
\begin{equation} 
P(0)=1+{\overline{u_t^{\rm s}}\over\omega(3-m)}
\end{equation} 
where $\overline{u_t^{\rm s}}=\tilde{u}_t^{\rm s}R/D^2=
\omega\big((4-3\omega)(m-3)+2(1-\omega)(4-2\omega-m)\big)/2$ 
(Hnatyk \cite{Hn87}), $\omega=2/(\gamma+1)$. \par 

Such an approach (substitution with 
$\tilde{r}^{-2}_{\rm s}\tilde{u}_t^{\rm s}$ instead of 
$\tilde{r}^{-2}\tilde{u}_t$ in (\ref{Eul-eq})) was also used by Laumbach 
\& Probstein (\cite{L-P}) to develop the sector approximation. \par

In PGA a power-law form for the pressure gradient 
\begin{equation} 
{dP(r)\over dr}=P^*_{\rm s}r^{l_{p,2}-1}\  
\end{equation} 
is used to give two-power approximation for the pressure 
\begin{equation} 
P(r)=P(0)+a_pr^{l_{p,2}}\ , 
\end{equation} 
where $a_p=P^*_{\rm s}/l_{p,2}$ and 
\begin{equation} 
\begin{array}{l}
{\displaystyle P(0)={(\gamma+1)^2(m_1-m)\over
3\gamma^2+20\gamma+1-(\gamma+1)(3\gamma+1)m}\ ,}\\ \\
{\displaystyle \hspace{0.4cm} l_{p,2}=
{3\gamma^2+20\gamma+1-(\gamma+1)(3\gamma+1)m 
\over 2(\gamma^2-1)} \ .}
\end{array}
\end{equation} 

\begin{figure}
\epsfxsize=8.6truecm
\centerline{\epsfbox{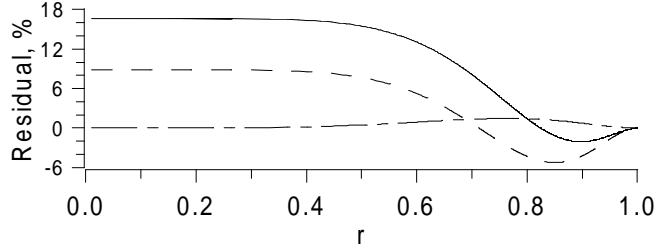}}
\caption[]{Relative differences of Ostriker \& McKee two-power 
approximation of the self-similar solution for the uniform medium. 
$\gamma=5/3$. 
Lines are the same as on Fig.~\ref{accuracy_Taylor_Kahn}. 
           }
\label{accuracy_Ostr-McKee}
\end{figure}

Accuracy in determination of $\alpha_A$ and $P(0)$ in approximations of 
Ostriker \& McKee is shown in table \ref{alpha_comp} and in revealing 
the flow parameters on Fig.~\ref{accuracy_Ostr-McKee}. \par

\subsection{Cavaliere \& Messina approximation of $\alpha_A$} 

Cavaliere \& Messina (\cite{Cavaliere-Messina-76}) with a simple  
technique approximate the equations for the radius and velocity 
of shock in the power-law medium (\ref{rho-power}) and 
$E_o(t)\propto t^{s}$. 
For $s=0$ his approximation gives 
\begin{equation} 
\label{beta-Cav-Mess} 
\beta_A={4\over \gamma^2-1}\left({\gamma-1\over\gamma+1}+{1\over 2}
{N+1-m\over N+1}\right).
\end{equation}

\subsection{Approximate methods for an explosion in medium with 
	arbitrary large-scale nonuniformity}

In this subsection we pointed out a number of approximate methods for 
description of a point explosion in arbitrary nonuniform medium. 
These methods may also be 
applicable for a medium with power-law density variation. 
Bisnovatyi-Kogan \& Silich (\cite{BK-Syl}) and Hnatyk (\cite{Hn87}) 
have given the reviews of these methods, their applications and accuracy. 

\subsubsection{Thin-layer approximation}

Thin-layer approximation is firstly introduced by Chernyi (\cite{Chernyi}) 
and used by Kompaneets (\cite{Komp}) and other authors to find analytical 
solutions for evolution of the shock front in a number of type of nonuniform 
media. It is assumed in this approach that all swept-up mass is concentrated 
in the infinitely thin layer just after shock front and the motion is 
stimulated with the hot gas inside the shocked region with uniform pressure 
distribution $P(r)=0.5$ (excepting $P_{\rm s}=1$). 
Layer of the gas moves with velocity $u_{\rm s}$. This method was developed 
to calculate anly the shock front dynamics and therefore does not allow to 
reveal the distribution of the fluid parameters behind the shock front. \par

Thin-layer approximation gives for spherical blastwave in the uniform medium 
(Andriankin et al. \cite{Andriankin-et-al}) 
\begin{equation} 
\label{alpha_A-TL} 
\alpha_A={16\pi(3\gamma-1)\over 75(\gamma-1)(\gamma+1)^2}\ .
\end{equation}

\subsubsection{Sector approximation}

In the sector approximation, the characteristics of an one-dimentional flow 
find as decompositions into series about the shock front. \par

Laumbach \& Probstein (\cite{L-P}) have proposed the sector approximation 
applying it to 
spherical blastwaves in a plane-stratified exponential medium. Authors   
use Lagrangian coordinate $a$ and 
propose to approximate pressure variation in the form equivalent to  
$P(a)=1+P_a^{\rm s}(a-1)$ (Hnatyk \cite{Hn87}). 
Density variation is given by the adiabaticity condition and relation 
$r=r(a)$ by continuity equation. Fluid velocity field is not determined. 
For shock radius and its velocity Laumbach \& Probstein approximation yeilds 
in the uniform medium limit 
\begin{equation} 
\label{alpha_A-SA} 
\alpha_A={32\pi(4\gamma^2-\gamma+3)\over 225(\gamma-1)(\gamma+1)^3}\ .
\end{equation}

Gaffet (\cite{Gaffet,Gaffet-81}) uses Lagrangian mass coordinates $\mu$ and 
finds pressure variation as a linear pressure approximation 
$P(\mu)=1+P_\mu^{\rm s}(\mu-1)$. 
Gaffet (\cite{Gaffet,Gaffet-81}) also propose to improve accuracy of 
the approximation, taking into account the second order coefficients 
in the series. 
Author calculates such coefficients in terms of Lagrangian mass 
coordinate $\mu$. Hnatyk (\cite{Hn87}), considering different 
modifications of the sector approximation, presents the coefficients 
up to the second order in terms of $a$. \par 

\begin{figure}
\epsfxsize=8.6truecm
\centerline{\epsfbox{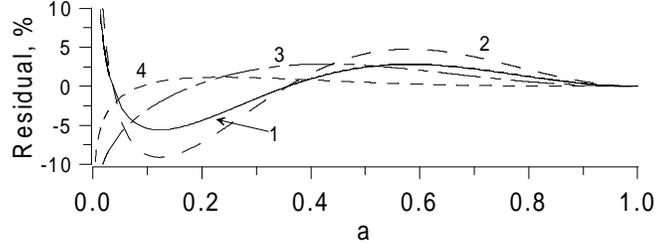}}
\caption[]{Accuracy of Hnatyk approximation of Sedov solution for the 
uniform medium. 
Lines: 1 -- $\rho(a)$, 2 -- $P(a)$, 3 -- $u(a)$, 4 -- $r(a)$. $\gamma=5/3$.
           }
\label{accuracy_Hn}
\end{figure}

\subsubsection{Hnatyk approximation}
\label{Hn-app}

Hnatyk (\cite{Hn87}) introduces also the idea to aproximate first\-ly 
the relation $\tilde{r}=\tilde{r}(a,t)$ between the Lagrangian $a$ and 
Eulerian $r$ coordinates of the gas element in each sector of shocked 
region. Density $\rho$, pressure $P$ and velocity $u$ variation behind 
the shock front are exactly deduced from this relation. Really, the 
continuity equation 
\begin{equation}
\tilde{\rho}^o(\tilde{a}) \tilde{a}^N d\tilde{a}=
\tilde{\rho}(\tilde{r}) \tilde{r}^N d\tilde{r}
\label{cont-eq-rho} 
\end{equation}
gives us the density distribution 
\begin{equation}
\label{ro(at)} 
\rho(a)={\tilde{\rho}(a,t)\over \tilde{\rho}^{\rm s}(t)}= 
{\tilde{\rho}^{o}(\tilde{a})\over \tilde{\rho}(R,t)}
	\left({\tilde{a}\over \tilde{r}(\tilde{a},t)}\right)^N
	\left({\partial \tilde{r}(\tilde{a},t)\over\partial \tilde{a}}
\right)^{-1}, 
\end{equation}
the equation of adiabaticity 
\begin{equation} 
\tilde{P}(\tilde{a},t)=K\tilde{\rho}(\tilde{a},t)^{\gamma}
\end{equation} 
yields the distribution of pressure
\begin{equation} 
\label{P(at)} 
P(a)=
{\tilde{P}(\tilde{a},t)\over \tilde{P}^{\rm s}(t)}=
	\biggl({\tilde{\rho}^{o}(\tilde{a})\over
	\tilde{\rho}^{o}(R)}\biggr)^{\!1-\gamma} \biggl({D(\tilde{a})\over
	D(R)}\biggr)^{\!2}\biggl({\tilde{\rho}(\tilde{a},t)\over \tilde{\rho}(R,t)} 
	\biggr)^{\!\gamma}
\end{equation} 
and relation $\tilde{r}=\tilde{r}(\tilde{a},t)$ gives velocity 
\begin{equation}
\label{u(at)}
u(a)=
{\tilde{u}(\tilde{a},t)\over \tilde{u}^{\rm s}(t)}=
{\gamma+1\over 2}{1\over D(R)}
	{d\tilde{r}(\tilde{a},t)\over dt}\ .
\end{equation}

Author propose to approximate $r(a)$ as 
\begin{equation}
r(a)=a^\alpha\exp\big(\beta(a-1)\big)\  
\label{r(a)-Hn}
\end{equation}
with 
\begin{equation}
\alpha=(r^{\rm s}_{a})^2-r^{\rm s}_{aa} \quad {\rm and}\quad  
\beta=r^{\rm s}_{aa}+r^{\rm s}_{a}-(r^{\rm s}_{a})^2 \ .
\end{equation}
Such an expression ensures the edge condition $r(0)=0$, $r_{\rm s}=1$ 
and values of the derivatives 
\begin{equation} 
r^{\rm s}_a = 1-\omega,
\label{ras}
\end{equation} 
\begin{equation} 
r^{\rm s}_{aa}=\omega(1-\omega) \bigl[ 3B+N(2-\omega)-m\bigr] 
\label{raas}
\end{equation} 
where $B=R\ddot R/\dot R^2$, $\dot R=dR/dt$ is the shock velocity, 
$m=-d\ln\rho^o(R)/d\ln R$, 
subscript "$a$" denotes a partial derivative in respect to $a$. \par

This approximation is accurate near the shock front, but around 
the explosion site (for $a<0.1$ or $r<0.4$) characteristics do not 
restore correctly (Fig.~\ref{accuracy_Hn}). 
This approximation does not take into consideration any derivatives of $r(a)$ 
near the center and the distributions of $\rho(a)$, $P(a)$, $u(a)$ do not 
bind there, causing such a situation. 
This approximation is extended to the central region in subsection 
\ref{app-Hn-imp}. \par 

\section{Extension of Taylor approximation to $m\neq 0$}
\label{Taylor-ext}

In this section Taylor's technic is applied to the case of a medium 
with the power-law density distribution (\ref{rho-power}) with $m<m_1$. 
Ostriker \& McKee 
(\cite{Ostriker-McKee-88}) give the coefficients in the approximation 
(\ref{app-Taylor-base}) for $u(r)$ in such a case. 
Approximated velocity variation is (\ref{Taylor-u}) with 
$n=P^*_{\rm s}-2=(7\gamma-1-m\gamma(\gamma+1))/(\gamma^2-1)$. 
Substitution with (\ref{app-Taylor-base}) into the equations of 
continuity and state gives 
\begin{equation}
{\rho_r\over\rho}={3+\alpha\gamma(n+2)r^{n-1}-m\gamma\over (\gamma-1)r-\alpha\gamma r^{n}}\ , 
\end{equation}
\begin{equation}
{P_r\over P}={\alpha\gamma^2(n+2)r^{n-1}\over (\gamma-1)r-\alpha\gamma r^{n}}\ .
\end{equation}
After integration, pressure variation will be expressed 
with (\ref{Taylor-T}) where  
\begin{equation}
q=
{2\gamma^2+7\gamma-3-m\gamma(\gamma+1)\over 7-\gamma-m(\gamma+1)} \ .
\end{equation}
Density is 
\begin{equation}
\rho(r)=
r^{\ (3-m\gamma)/(\gamma-1)}\ \left({\gamma+1\over\gamma}-{r^{n-1}\over\gamma}\right)^{-p},
\end{equation}
where 
\begin{equation}
p=
{2\big(\gamma+5-m(\gamma+1)\big)\over 7-\gamma-m(\gamma+1)} \ .
\end{equation}
Eq.~(\ref{Taylor-a}) gives $a(r)$ with 
\begin{equation}
s=
{\gamma+1\over 7-\gamma-m(\gamma+1)} \ .
\end{equation}

\begin{figure}
\epsfxsize=8.6truecm
\centerline{\epsfbox{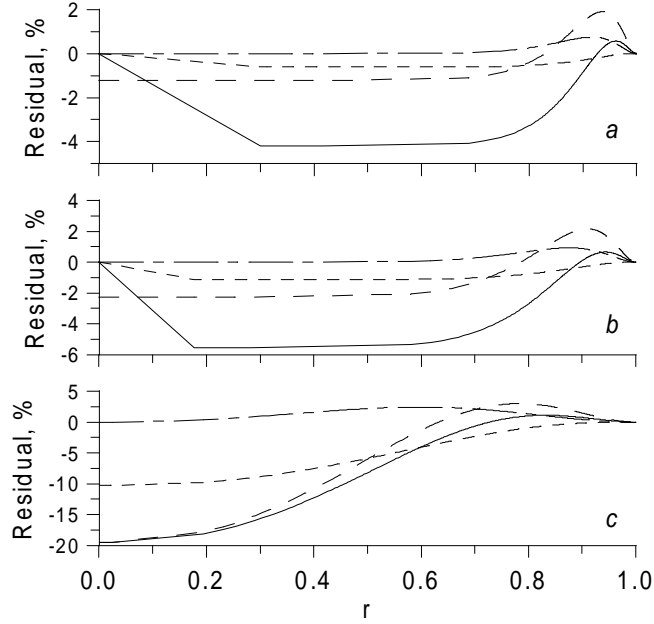}}
\caption[]{{\bf a-c.} Accuracy of the extended Taylor approximation of the 
self-similar solution in the power-law medium (\ref{rho-power}):  
{\bf a}~relative differences of the approximation for $m=-4$, 
{\bf b}~relative differences for $m=-2$, 
{\bf c}~relative differences for $m=1$. 
Lines are the same as on Fig.~\ref{accuracy_Taylor_Kahn}. $\gamma=5/3$.
           }
\label{accuracy_Taylor-ext_N-2}
\end{figure}

Exponents $n=6-5m/2$, $q=(16-5m)/\big(3(2-m)\big)$, $p=(5-2m)/(2-m)$ and 
$s=1/(2-m)$ for $\gamma=5/3$. 
This extended Taylor approximation is compared with the exact solution on 
Fig.~\ref{accuracy_Taylor-ext_N-2}. 

\section{Approximations of the Sedov solution 
	in Lagrangian coordinates}

In this section we present two analytical approximations of the self-similar 
solution for a medium with the power-law density distribution 
expressed in Lagrangian geometric coordinates $a$. 

\subsection{Flow characteristic distributions} 

Exact expressions for normalized density $\rho$ and pressure $P$ 
variations behind the shock front moving into the power-law medium 
(\ref{rho-power}) follow from (\ref{ro(at)}) and (\ref{P(at)}): 
\begin{equation}
{\displaystyle 
\rho(a)=
{\gamma-1\over\gamma+1}\cdot a^{N-m}\cdot \left(r(a)^N
\cdot r_a(a)\right)^{-1} }\ , 
\label{appr_3a}
\end{equation}
\begin{equation}
{\displaystyle 
P(a)=
\left({\gamma-1\over\gamma+1}\right)^{\gamma}\cdot 
	a^{N(\gamma-1)-1}\cdot \left(r(a)^N
	\cdot r_a(a)\right)^{-\gamma} }\ .
\label{appr_3b}
\end{equation}

Distribution of the fluid velocity $u(a)$ may be found from (\ref{u(at)}). 
Due to $\tilde{r}=rR$ time derivative $d\tilde{r}/dt=Rr_t+Rr_aa_t+rD$ 
(subscript "$t$" denotes a partial derivative in respect to $t$). 
We have also that $a_t=-aD/R$ and, in the self-similar case, $r_t=0$. So, 
\begin{equation}
u(a)={\gamma+1\over 2}\Bigl(r(a)-r_a(a)a\Bigr)\ .
\label{u-distr-r^w}
\end{equation}

The distribution $\mu(a)$ follows from the definition (\ref{mu-def}) and 
(\ref{cont-eq-rho}): 
\begin{equation} 
\mu(a)=a^{(N+1)-m}\ .
\label{mu(a)}
\end{equation}

\subsection{Self-similar constant $\alpha_A$}
\label{alpha-A}

Self-similar constant $\alpha_A(N,\gamma,m)$ in equations for $R$ and 
$D$ (\ref{R_s_Sedov-ro^w})-(\ref{D_Sedov-ro^w}) obtains from 
(\ref{alpha_A-vs-beta_A}) and (\ref{beta_A}): 
\begin{equation} 
\label{alpha_A} 
\alpha_A={8\over\gamma^2-1}\cdot{\sigma\over(3+N-m)^2}\cdot
\left(I_{\rm K}+I_{\rm T}\right)\ ,
\end{equation}
with
\begin{equation}
I_{\rm K}
={\gamma^2-1\over 4}
\int\limits_0^1 \Bigl(r(a)-r_a(a) a\Bigr)^2a^{N-m}da\ ,  \quad 
I_{\rm T}
=\left({\gamma-1\over \gamma+1}\right)^\gamma
\int\limits_0^1 \Bigl(r(a)^Nr_a(a)\Bigr)^{1-\gamma}a^{N(\gamma-1)-1}da\ . 
\end{equation}

\subsection{Factor $C$ and exponent $x$}
\label{factor-C}

In Sedov self-similar solution, if $r\rightarrow 0$ then the dependence 
$r(a)$ is 
\begin{equation} 
r=C\cdot a^x. 
\label{r_Ca^x}
\end{equation} 

For $m<m_1$, if we substitute (\ref{r_Ca^x}) 
into (\ref{appr_3b}) we obtain the connection between the factor $C$ and  
normalized central pressure $P(0)$: 
\begin{equation} 
\label{C-Sedov-2}
C= \left({\gamma\over\gamma+1} \cdot 
   P(0)^{-1/\gamma}\right)^{1/(N+1)}.  
\end{equation}
We have to put $x=(\gamma-1)/\gamma$ during this transformation 
in order to satisfy condition $P(0)\neq 0$. 
In the case $m=m_1$ the exact solution (\ref{solut-m_1}) gives $x$ and 
$C=1$. 
General formula for exponent $x$ is 
\begin{equation} 
x=\left\{\matrix {(\gamma-1)/\gamma     & {\rm for}\ m<m_1 &\cr
		  (\gamma-1)/(\gamma+1) & {\rm for}\ m=m_1 &.\cr}\right.
\label{exponent-x}
\end{equation} 

Analytical expressions for $P(0)$ from self-similar solution are presented 
in Appendix \ref{app-2}. Calculated values of $P(0)$ and $C$ for 
a number of $N$, $\gamma$ and $m$ are shown in table~\ref{C-value}. \par 

\begin{table}[t]
\caption[]{ $P(0)$ calculated according to the self-similar 
	solution and $C$ for a strong point explosion 
	in the power-law medium. \\  } 
\begin{center}
\begin{tabular}{cccccccc}
\hline 
\noalign{\smallskip}
N&m&&\multicolumn{2}{c}{$P(0)$}&&\multicolumn{2}{c}{$C_{\rm A}$}\\
\noalign{\smallskip}
\cline{4-5}\cline{7-8}
\noalign{\smallskip}
&&&$\gamma=7/5$&$\gamma=5/3$&&$\gamma=7/5$&$\gamma=5/3$\\
\noalign{\smallskip}
\hline
\noalign{\smallskip}
0&0&&0.3900&0.3532&&1.1429&1.1670\\
\noalign{\smallskip}
1&0&&0.3729&0.3215&&1.0863&1.1112\\
\noalign{\smallskip}
2&0&&0.3655&0.3062&&1.0618&1.0833\\
\noalign{\smallskip}
&-4&&0.4268&0.3954&&1.0233&1.0293\\
\noalign{\smallskip}
&-3&&0.4193&0.3848&&1.0276&1.0350\\
\noalign{\smallskip}
&-2&&0.4088&0.3696&&1.0339&1.0433\\
\noalign{\smallskip}
&-1&&0.3928&0.3463&&1.0438&1.0570\\
\noalign{\smallskip}
& 1&&0.3087&0.2217&&1.1054&1.1556\\
\noalign{\smallskip}
& 2&&0.1273&0.0000&&1.3648&1.0000\\
\noalign{\smallskip}
\hline
\end{tabular}
\end{center}
\label{C-value}
\end{table}

\begin{table}
\caption[]{Derivatives of the relation between Lagrangian and 
	Eulerian coordinates at shock front moving into the power-law 
	medium (\ref{rho-power}).\\ } 
\begin{center}
\begin{tabular}{l}
\hline
\noalign{\smallskip}
Derivative \hfill{$\gamma$}\\
\noalign{\smallskip}
\hline
\noalign{\medskip}
\hfill{$\gamma=7/5$}\\
\noalign{\medskip}
${\displaystyle r_a^{\rm s}={1\over 6}}$\\
\noalign{\medskip}
${\displaystyle r_{aa}^{\rm s}={5\over 2^3 3^3}\left(-2N+3m-9\right)}$\\
\noalign{\medskip}
${\displaystyle r_{aaa}^{\rm s}={5\over 2^5 3^5}\left(-2N^2+9Nm+183N-9m^2-270m+675\right)}$\\
\noalign{\medskip}
\hline
\noalign{\medskip}
\hfill{$\gamma=5/3$}\\
\noalign{\medskip}
${\displaystyle r_a^{\rm s}={1\over 4}}$\\
\noalign{\medskip}
${\displaystyle r_{aa}^{\rm s}={3\over 2^6}\left(-N+2m-6\right)}$\\
\noalign{\medskip}
${\displaystyle r_{aaa}^{\rm s}={3\over 2^9}\left(Nm+19N-2m^2-36m+94\right)}$\\
\noalign{\medskip}
\hline
\end{tabular}
\end{center}
\label{r^s_deriv}
\end{table}

\subsection{Derivatives at shock front} 

Expressions for the derivatives $r_a^{\rm s}$, $r_{aa}^{\rm s}$, 
$r_{aaa}^{\rm s}$ may be obtained 
with the technic of Gaffet (\cite{Gaffet}) from the set of  
hydrodynamic equations for perfect gas and conditions on the shock front 
(see Hnatyk \& Petruk (\cite{Hn-Pet-99}) for details). 
Derivatives $r_a$, $r_{aa}$ are given with (\ref{ras})-(\ref{raas}) and  
\begin{equation} 
\begin{array}{l} 
r^{\rm s}_{aaa}=\omega
(1-\omega)\Bigl[ 3(7-5\omega)B^2+ \\ \qquad\quad +\bigl[
(-5\omega^2+4\omega +8)N+(4\omega -11)m\bigr]B+  \\ \qquad\quad
+\omega(2\omega ^2-7\omega+6)N^2+(\omega^2+\omega -4)Nm- \\ \qquad\quad
-\omega(2-\omega)N-(\omega-2)m^2+(2\omega-1)m+           \\ \qquad\quad
+(2\omega-1)m'+(6\omega-4)Q \Bigr],
\label{raaas}
\end{array} 
\end{equation} 
where $Q=R^2R^{(3)}/\dot R^3$ and $m'=-dm/d\ln R$. \par

In the power-law medium (\ref{rho-power}) $m'=0$. Taking into consideration 
the equations for the shock radius (\ref{R_s_Sedov-ro^w}) and shock velocity 
(\ref{D_Sedov-ro^w}) we may also write 
\begin{equation}
B=-{N-m+1\over2}, \ \ Q={(N-m+1)(N-m+2)\over2}. 
\end{equation}
Reduced expressions for the derivatives $r_a^s$, $r_{aa}^s$, $r_{aaa}^s$ 
are shown in table~\ref{r^s_deriv}. 

\begin{figure}[t]
\epsfxsize=8.6truecm
\centerline{\epsfbox{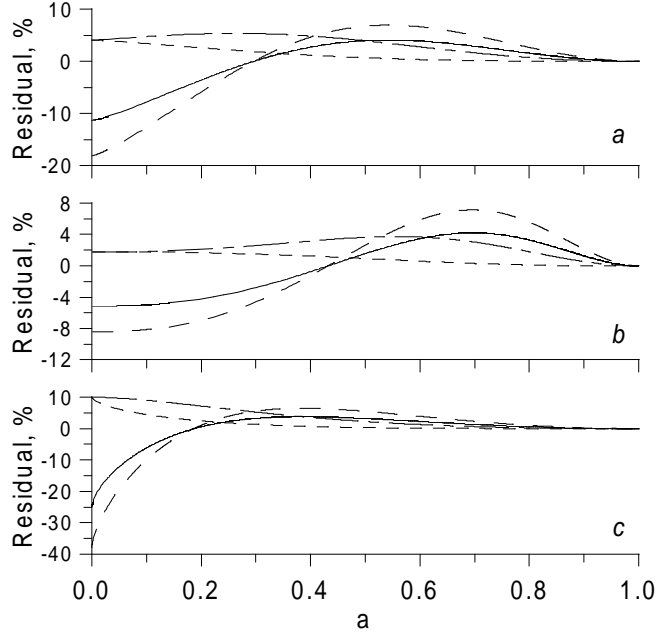}}
\caption[]{{\bf a-c.} Accuracy of the second order approximation of the 
self-similar solution in the power-law medium (\ref{rho-power}) for 
$\gamma=5/3$ and $N=2$:   
{\bf a}~relative differences for $m=0$, 
{\bf b}~relative differences for $m=-2$, 
{\bf c}~relative differences for $m=1$. 
Lines are the same as on Fig.~\ref{accuracy_Hn}.
           }
\label{accuracy-Hn-imp}
\end{figure}

\subsection{Second order approximation}
\label{app-Hn-imp}

So, to approximate the self-similar solution, we approximate the relation 
$r=r(a)$ between 
Eulerian $r$ and Lagrangian $a$ coordinates of flow elements. 
Following to Hna\-tyk's approach (\ref{r(a)-Hn}) and like to 
relation (\ref{Kahn-a}), $r=r(a)$ may be aproximated in the form 
\begin{equation}
r(a)=
a^x\exp\Big(\alpha (a^{\beta}-1)\Big)\  
\label{Hn-impruved}
\end{equation}
with $x$ given by (\ref{exponent-x}) and 
\begin{equation}
\alpha={(r^{\rm s}_{a}-x)^2\over
	r^{\rm s}_{aa}+r^{\rm s}_{a}-(r^{\rm s}_{a})^2}\ ,\quad 
\beta={r^{\rm s}_{aa}+r^{\rm s}_{a}-(r^{\rm s}_{a})^2
	\over r^{\rm s}_{a}-x}\ ,
\end{equation}
or, after substitution with (\ref{ras})-(\ref{raas}), 
\begin{equation}
\begin{array}{l}
{\displaystyle \alpha={2(1-\omega-x)^2\over\omega(1-\omega)(N+m-1-2N\omega)}\ ,}\\ \\
{\displaystyle \beta=\alpha^{-1}(1-\omega-x) .}
\end{array}
\end{equation}
Such a second order approximation, besides $r(0)=0$, $r_{\rm s}=1$, 
$r^{\rm s}_{a}$, $r^{\rm s}_{aa}$, gives 
$(\partial\ln r/\partial\ln a)^{0}=x$, 
and, contrary to Hnatyk approximation, extends description of a flow to 
the central region. \par

Variations of $\rho(a)$, $P(a)$ and $u(a)$ follow from 
(\ref{appr_3a})-(\ref{u-distr-r^w}). For case $N=2$, $\gamma=5/3$ and $m<2$ 
these relations give $\beta=5(2-m)/8$ and 
\begin{equation} 
r(a)=a^{2/5}\exp\left(-{6\over25(2-m)}(a^{\beta}-1)\right)\!,
\label{r(a)-twoa}
\end{equation}
\begin{equation} 
\begin{array}{l} 
{\displaystyle \rho(a)=
\left({8\over5}-{3\over5}a^{\beta}\right)^{-1}\cdot a^{(9-5m)/5} }\\ \\ 
{\displaystyle \qquad\qquad\qquad\!\!
\times
\exp\left({18\over25(2-m)}(a^{\beta}-1)\right)\!, }
\end{array} 
\label{rho(a)-twoa}
\end{equation}
\begin{equation} 
{\displaystyle P(a)=\left({8\over5}-{3\over5}a^{\beta}\right)^{-5/3} 
\exp\left({6\over5(2-m)}(a^{\beta}-1)\right)\!, }
\label{P(a)-twoa}
\end{equation}
\begin{equation} 
u(a)=a^{2/5}\left({4\over5}+{1\over5}a^{\beta}\right)
\exp\left(-{6\over25(2-m)}(a^{\beta}-1)\right)\!.
\label{u(a)-twoa}
\end{equation}
Approximation (\ref{r(a)-twoa})-(\ref{u(a)-twoa}) may be considered as an 
inversion of Cox \& Franko approximation (\ref{Kahn-n})-(\ref{Kahn-a}). 
Unfortunately, accuracy of presented formulae is lower 
(Fig.~\ref{accuracy-Hn-imp}, table \ref{alpha_comp}). 

\subsection{Third order approximation}

In order to improve accuracy, we postulate the approximation $r=r(a)$ 
to give exact values of two additional derivatives: third order 
$r^{\rm s}_{aaa}$ and 
$(\partial r/\partial(a^x))^{0}=C$. Consideration of $r^{\rm s}_{aaa}$ is 
equivalent to consideration of the second order derivatives 
$\rho_{aa}^{\rm s}$, $P_{aa}^{\rm s}$, $u_{aa}^{\rm s}$ in expansion of 
relevant characteristics into the series near the shock front. 
This approximation is the same as used in the approximate hydrodynamical 
method 
for modelling the asymmetrical strong point explosion in the medium with a 
large-scale density nonuniformity (Hnatyk \& Petruk \cite{Hn-Pet-99}). 
Contrary to the method, we take here that both the self-similar 
constant $\alpha_A$ and factor $C$ are different for different $m$. \par

Namely, if at time $t$ the shock position is $R(t)$, we approximate a 
connection $r=r(a)$ as follows 
\begin{equation} 
r(a)=a^x\cdot 
(1+\alpha\cdot \xi + \beta \cdot \xi^2 + \varsigma \cdot \xi^3 + \delta \cdot 
\xi^4), 
\label{r(a)_approx}
\end{equation} 
where $\xi=1-a$. Coefficients $\alpha, \beta, \varsigma, \delta$ and exponent 
$x$ are choosen from the condition that the partial derivatives 
$r^{s}_{a}$, $r^{\rm s}_{aa}$, $r^{\rm s}_{aaa}$ 
at the shock front ($a=1$) as well as 
$(\partial\ln r/\partial\ln a)^{0}=x$ and $(\partial r/\partial (a^x))^{0}=C$ 
in the place of explosion ($a=0$) equal to their exact values: 
\begin{equation} 
\begin{array}{l} 
{\displaystyle 
\alpha = -r^{\rm s}_a +x, }\\ [0.2cm]
{\displaystyle 
\beta ={1\over 2}\cdot \bigl( 
r^{\rm s}_{aa}-2x\cdot r^{\rm s}_a + x(x+1)\bigr), }\\ [0.3cm]
\varsigma = {\displaystyle {1\over 6}\cdot } 
{\displaystyle 
\bigl(- r^{\rm s}_{aaa} +3x\cdot r^{\rm s}_{aa}- }\\ [0.3cm]
\quad\qquad -3x(1+x) \cdot r^{\rm s}_a+x(x+1)(x+2)\bigr), \\ [0.2cm]
{\displaystyle 
 \delta = C - (1 + \alpha + \beta + \varsigma).  }\\ \end{array} 
\!\!\!\!\!\!\!\!\!\!\!\!\!\!\!\!\!\!\!\!\!\!\!\!\!\!\!\!  
\end{equation} 
In terms of $a$ relation (\ref{r(a)_approx}) and its 
first derivative are 
\begin{equation}
r(a)=a^x(B_0-B_1 a+B_2 a^2-B_3 a^3+B_4 a^4),  \\ \\
\label{appr_2a}
\end{equation}
\begin{equation}
r_a(a)=a^{x-1}(A_0-A_1 a+A_2 a^2-A_3 a^3+A_4 a^4), 
\label{appr_2b}
\end{equation}
with \par
\begin{center}
\begin{tabular}{lll}
\noalign{\medskip}
$B_0=1+\alpha+\beta+\varsigma+\delta=C,$&&$A_0=xB_0,$    \\
\noalign{\smallskip}
$B_1=\alpha+2\beta+3\varsigma+4\delta,$ &&$A_1=(1+x)B_1,$\\
\noalign{\smallskip}
$B_2=\beta+3\varsigma+6\delta,$         &&$A_2=(2+x)B_2,$\\
\noalign{\smallskip}
$B_3=\varsigma+4\delta,$                &&$A_3=(3+x)B_3,$\\
\noalign{\smallskip}
$B_4=\delta,$                        &&$A_4=(4+x)B_4.$\\
\noalign{\medskip}
\end{tabular}
\end{center}

\begin{table}
\caption[]{Self-similar constant $\alpha_{\rm A}(N,\gamma,m)$ calculated 
	with third order approximation of $r(a)$ (\ref{r(a)_approx}). \\
          } 
\begin{center}
\begin{tabular}{crcc}
\hline 
\noalign{\smallskip}
N&m&\multicolumn{2}{c}{$\alpha_{\rm A}$}\\
\noalign{\smallskip}
\cline{3-4}
\noalign{\smallskip}
&&$\gamma=7/5$&$\gamma=5/3$\\
\noalign{\smallskip}
\hline
\noalign{\smallskip}
0& 0&1.0763&0.6018 \\
\noalign{\smallskip}
1& 0&0.9841&0.5644 \\
\noalign{\smallskip}
2& 0&0.8519&0.4944 \\
\noalign{\smallskip}
&-4&0.2295 &0.1270 \\
\noalign{\smallskip}
&-3&0.2960 &0.1650 \\
\noalign{\smallskip}
&-2&0.3966 &0.2232 \\
\noalign{\smallskip}
&-1&0.5598 &0.3192 \\
\noalign{\smallskip}
& 1&1.4631 &0.8722 \\
\noalign{\smallskip}
& 2&3.3537 &1.8235 \\
\noalign{\smallskip}
\hline
\end{tabular}
\end{center}
\label{alpha_A-calc}
\end{table}

\begin{table}[t]
\caption[]{Coefficients in approximation (\ref{appr_2a}). $\gamma=5/3$.  } 
\begin{center}
\begin{tabular}{rrllllll}
\hline
\noalign{\smallskip}
$N$&$m$&$B_0$&$B_1$&$B_2$&$B_3$&$B_4$\\
\noalign{\smallskip}
\hline
\noalign{\smallskip}
0& 0&1.1670     &0.1333     &-0.1127    &-0.1074    &-0.02833 \\
\noalign{\smallskip}
1& 0&1.1112     &-0.01510   &-0.24655   &-0.1530    &-0.03276 \\
\noalign{\smallskip}
2& 0&1.0833     &-0.05189   &-0.2130    &-0.08708   &-0.009294 \\
\noalign{\smallskip}
 &-4&1.0293     &0.2837     &0.9302     &0.9766     &0.3008   \\
\noalign{\smallskip}
 &-3&1.0350     &0.1744     &0.6152     &0.6971     &0.2213   \\
\noalign{\smallskip}
 &-2&1.0433     &0.07170    &0.3041     &0.4162     &0.1406   \\
\noalign{\smallskip}
 &-1&1.0570     &-0.01334   &0.01359    &0.1452     &0.06128  \\
\noalign{\smallskip}
 & 1&1.1556     &0.08965    &-0.1753    &-0.1471    &-0.03778 \\
\noalign{\smallskip}
 & 2&1          &0          &0          &0          &0        \\
\noalign{\smallskip}
\hline
\end{tabular}
\end{center}
\label{tabl_appr_1}
\end{table}

Distribution of $\rho(a)$, $P(a)$ and $u(a)$ obtain from 
(\ref{appr_3a})-(\ref{u-distr-r^w}). 
Self-similar constant $\alpha_{\rm A}$ is given with (\ref{alpha_A}). 
To simplify the procedure, 
numerical values of $\alpha_A(N,\gamma,m)$ in this appoximation 
are presented in table~\ref{alpha_A-calc}.  
Table~\ref{tabl_appr_1} gives ready-calculated values of the coefficients 
in the approximation (\ref{appr_2a}) for a number of cases. 
\par

\begin{figure}
\epsfxsize=8.8truecm
\centerline{\epsfbox{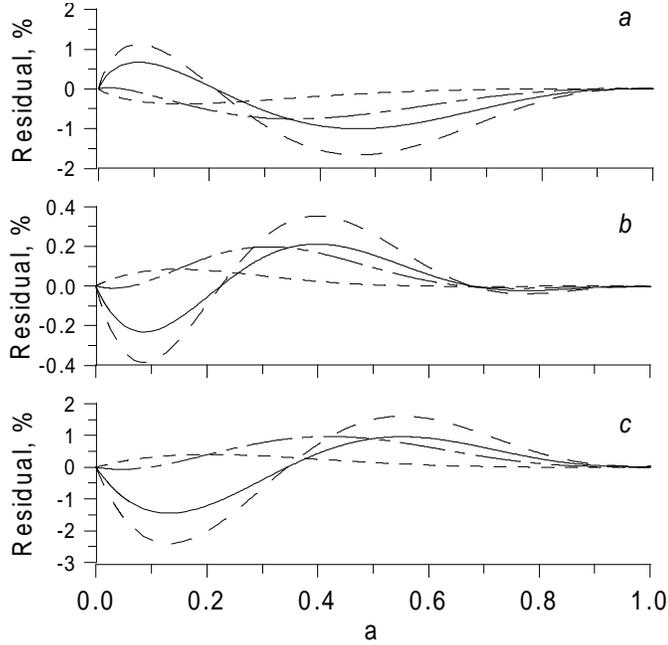}} 
\caption[]{{\bf a-c.} Accuracy of the third order approximation of the Sedov 
solution in the uniform medium ($m=0$) for $\gamma=5/3$: 
{\bf a}~relative differences of the approximation for $N=0$, 
{\bf b}~relative differences for $N=1$, 
{\bf c}~relative differences for $N=2$. 
Lines are the same as on Fig.~\ref{accuracy_Hn}.
           }
\label{O(m0)}
\end{figure}

\begin{figure}[t]
\epsfxsize=8.8truecm
\centerline{\epsfbox{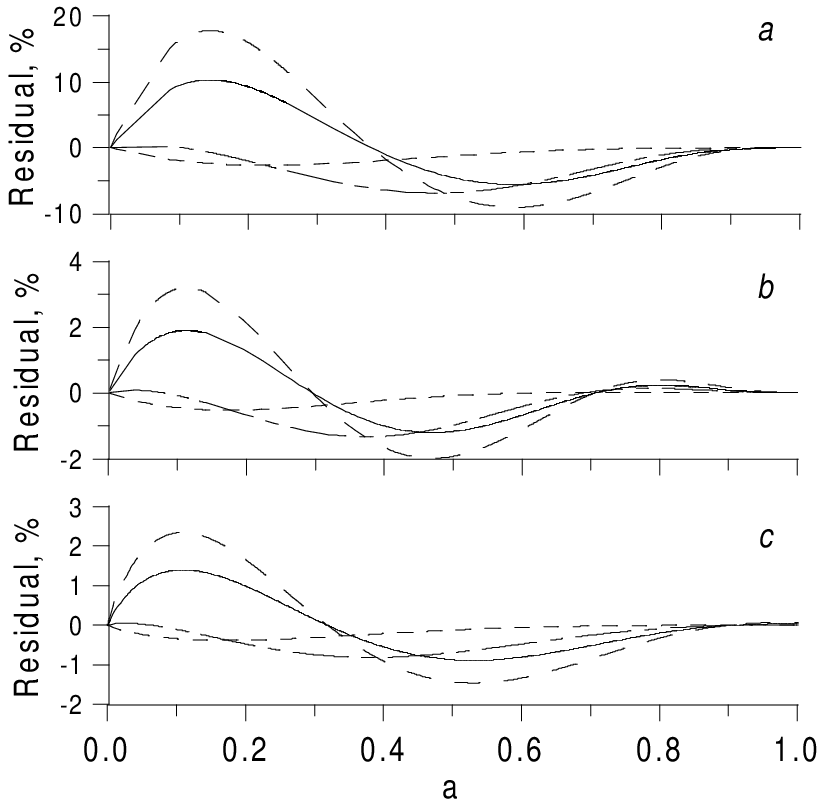}} 
\caption[]{{\bf a-c.} Accuracy of the third order approximation 
for power-law medium, $\gamma=5/3$ and $N=2$:  
{\bf a}~relative differences for $m=-4$, 
{\bf b}~relative differences for $m=-2$, 
{\bf c}~relative differences for $m=1$. 
Lines are the same as on Fig.~\ref{accuracy_Hn}.
           }
\label{O(N2)}
\end{figure}

The accuracy of flow characteristic distributions in this approximation 
is high for uniform medium (Fig.~\ref{O(m0)}). Approximation coinsides 
with the exact solution (\ref{solut-m_1}) for case $m=m_1$. 
For other $m\neq0$, differences increase with increasing $|m|$ 
but maximal errors reveal in the region with low densities (Fig.~\ref{O(N2)}).  
We compare also numerical values of $\alpha_A$ and $P(0)$ in this 
approximation with those from exact Sedov solution in table~\ref{alpha_comp}. 
$\alpha_A$ in the approximation is close to the exact values and gives 
accurate shock radius $R$ and velocity $D$. \par


\section{Conclusions}

In this paper, we review approximations of the self-similar 
solution for a strong point explosion in the power law medium 
$\rho^o\propto r^{-m}$ and compare their accuracy with the exact Sedov 
solution of the problem. Different approaches found on the different basic 
approximations. Namely, Taylor (\cite{Taylor-50}) and Ostriker \& McKee 
(\cite{Ostriker-McKee-88}) approximate first\-ly the fluid 
velocity variation behind the shock front. Taylor used approximated 
$u(r)$ substituting it into the hydrodynamic equations to obtain 
full description of the flow. Contrary to this, Ostriker \& McKee 
approximate $\rho(r)$ and $P(r)$ independently. Kahn 
(\cite{Kahn}) technic, used also by Cox \& Franco (\cite{Cox_Fanko-81}), 
consists in approximation of the fluid mass variation $\mu(r)$ and further 
usage of the system of hydrodynamic equations. Gaffet (\cite{Gaffet}), 
Laumbach \& Probstein (\cite{L-P}), 
Ostriker \& McKee (\cite{Ostriker-McKee-88}) base their approaches 
on the approximation of $P(\mu)$ or $P(r)$. Thin layer approximation may also 
be included into this group. Hnatyk (\cite{Hn87}) take approximation 
of the connection between Eulerian and Lagran\-gi\-an coordinates as basic 
relation. So, practically all possible approaches are used 
to have approximation for the self-similar solution. \par

In this paper we apply Taylor's methodology to discribe a strong point 
explosion in the power-law medium, extending his approximation written for 
uniform medium, and write also two approximations expressed in Lagran\-gi\-an 
geometric coordinates, approaching $r(a)$ with different accuracy. \par

Errors of all approximations are caused only by errors in the basic 
approximation. When the first approximation has higher accuracy we 
have more accurate approximation for parameters of the shock and flow. \par

\begin{table*}
\caption[]{ Comparision of the self-similar constant 
$\alpha_{\rm A}$ and pressure $P(0)$ calculated with: 
S --  Sedov (\cite{Sedov-1946b}) solution (Kestenboim et al. \cite{Kestenb}); 
T --  Taylor (\cite{Taylor-50}) approximation;
CF -- approximation of Cox \& Franco (\cite{Cox_Fanko-81}); 
LVA, OPA and TPA of Ostriker \& McKee (\cite{Ostriker-McKee-88}); 
CM -- approximation of Cavaliere \& Messina (\cite{Cavaliere-Messina-76}); 
TL -- thin-layer (\ref{alpha_A-TL}) approximation; 
LP -- approximation (\ref{alpha_A-SA}) of Laumbach \& Probstein (\cite{L-P}); 
SOA -- second order (\ref{Hn-impruved}) and 
TOA -- third order (\ref{r(a)_approx}) approximations. 
Uniform medium, $\gamma=5/3$ and $N=2$. \\
          } 
\begin{center}
\begin{tabular}{lcccccccccccccc}
\hline
\noalign{\smallskip}
&S&T&CF&LVA&OPA/LPA&TPA/PGA&CM&TL&LP&SOA&TOA\\
\noalign{\smallskip}
\hline
\noalign{\smallskip}
$\alpha_A$&0.4936 &0.4957 &0.4930 &0.5386 &0.5027 &0.4957  &0.5655 &0.5655 &0.4398 &0.4981  &0.4944 \\
\noalign{\smallskip}
$P(0)$	  &0.3062 &0.2855 &0.3140 &--     &0.3333 &0.3333  &--     &0.5000 &0.3333 &0.2507  &0.3062 \\
\noalign{\smallskip}
\hline
\end{tabular}
\end{center}
\label{alpha_comp}
\end{table*}


\appendix

\section*{Appendix: central pressure $P(0)$}
\label{app-2}

In this appendix, we give exact expression for $P(0)$ in self-similar 
solution when $m\leq m_1$ (Sedov \cite{Sedov}) and when $m=m_2$ 
(Korobejnikov \& Rjazanov \cite{Korobejnikov-Rjazanov-59}). 
These relations complite the full set of formulae to build the third order 
approximation of the Sedov solution for any $\gamma$, 
$m\le \min(N+1,m_1)$ and type of symmetry (plane, cylindrical or spherical 
blastwave). \par

$P(0)=0$ for $m=m_1$. \par

In the case of $m<m_1$ and $m\neq m_2$ 
\begin{equation} 
P(0)=\left(1\over2\right)^{\varepsilon_1}
\left(\gamma+1\over\gamma\right)^{\varepsilon_2}
\left(m-m_3\over m-m_1\right)^{\varepsilon_3\varepsilon_4}\ ,
\end{equation}
$$
\begin{array}{l}
\varepsilon_1=\displaystyle{ {2(N+1)\over N+3-m}\ ,}\\ \\
\varepsilon_2=\displaystyle{ {2(N+1)\over N+3-m}-{\gamma\big(N+1-m\big)\over(N+1)(2-\gamma)-m}\ ,}\\ \\
\varepsilon_3=\displaystyle{ {(N+1-m)(N+3-m)\over(N+1)(2-\gamma)-m}+m-2\ ,}\\ \\
\varepsilon_4=\displaystyle{ {\gamma+1\over(N+1)(\gamma-1)+2}-{2\over N+3-m} \hspace{1.9cm}}\\ \\
\hspace{3.85cm}     \displaystyle{ +{\gamma-1\over\gamma(2-m)+N-1}\ .}
\end{array}
$$
If $m=m_2$ then 
\begin{equation} 
P(0)=\left(1\over2\right)^{\varepsilon}
\left(\gamma+1\over\gamma\right)^{\varepsilon\gamma N/(N+1)}
\exp\left(-{\gamma\over 2}\ \varepsilon\right)\ ,
\end{equation}
$$
\varepsilon=\displaystyle{ {2(N+1)\over (N+1)(\gamma-1)+2}
\ .}\hspace{4.08cm}
$$



\end{document}